\documentclass[doublespacing]{elsart}

\usepackage{natbib}
\usepackage{float}
\usepackage{subfigure}
\usepackage{rotating}
\usepackage{longtable}
\usepackage{amsmath}

\usepackage{times}

\usepackage{graphicx}

\usepackage{amssymb}

\usepackage{lineno}

\linenumbers

\journal{Icarus}

\begin{document}

\newcommand \farcs{\mbox{$.\!\!^{\prime\prime}$}}%
\let\farcs\farcs

\begin{frontmatter}

\title{Discovery of a young asteroid cluster associated with P/2012~F5 (Gibbs)}

\author[bg]{Bojan Novakovi\'c}
\ead{bojan@matf.bg.ac.rs}
\author[hi]{Henry H. Hsieh}
\author[to]{Alberto Cellino}
\author[hi]{Marco Micheli}
\author[tng]{Marco Pedani}

\address[bg]{Department of Astronomy, Faculty of Mathematics, University of Belgrade,
Studentski trg 16, 11000 Belgrade, Serbia}
\address[hi]{Institute for Astronomy, University of Hawaii, 2680 Woodlawn
Drive, Honolulu, HI 96822, USA}
\address[to]{INAF--Osservatorio Astrofisico di Torino, Via Osservatorio 20, I-10025
Pino Torinese, Italy}
\address[tng]{Fundaci\' on Galileo Galilei - INAF Rambla Jos\' e Ana Fern\' andez P\' erez, 7
38712 Bre\~ na Baja, TF - Spain }

\begin{abstract}
We present the results of our search for a dynamical family around
the active asteroid P/2012~F5 (Gibbs). By applying the
hierarchical clustering method, we discover an extremely compact
9-body cluster associated with P/2012~F5. The statistical
significance of this newly discovered Gibbs cluster is estimated to be $>99.9$\%,
strongly suggesting that its members share a
common origin. The cluster is located in a dynamically cold region of the outer main-belt at a
proper semi-major axis of $\sim$3.005~AU, and all members are found to be dynamically stable over
very long timescales. Backward numerical orbital integrations
show that the age of the cluster is only $1.5\pm0.1$~Myr.
Taxonomic classifications are unavailable for most of the
cluster members, but SDSS spectrophotometry available for two cluster members
indicate that both appear to be $Q$-type objects.  We also estimate a lower limit of the size of the
parent body to be about 10~km, and find that the impact event which produced
the Gibbs cluster is intermediate between a cratering and a catastrophic
collision. In addition, we search for new main-belt comets in
the region of the Gibbs cluster by observing seven asteroids either
belonging to the cluster, or being very close in the space of
orbital proper elements. However, we do not detect any convincing
evidence of the presence of a tail or coma in any our targets. Finally, 
we obtain optical images of P/2012~F5, and find absolute $R$-band and $V$-band magnitudes
of $H_R=17.0\pm0.1$~mag and $H_V=17.4\pm0.1$~mag, respectively, corresponding to
an upper limit on the diameter of the P/2012~F5 nucleus of $\sim$2 km.
\end{abstract}


\end{frontmatter}

\section{Introduction}
\label{s:intro}

Asteroid families are believed to originate from catastrophic
fragmentations of single parent bodies \citep{zappala2002}. They are
very useful for studying various open problems in asteroid science
\citep{2010LNP}, and have been extensively investigated for
almost a century. In principle, it is clear that ``fresh'' young families,
only slightly evolved since the epoch of their formation, may
provide more direct information about the collisional events from
which they originated. In the last decade, our knowledge about such young
families has been increased significantly. Several new ones have
been discovered \citep[e.g.,][]{karin2002,nes2003,
datura2006,nes2006,nes2008,pravok2009,nov2010theo,vok2011,nov2012a,nov2012lorre}, and
many have been the subjects of detailed investigations
\citep[e.g.,][]{vernazza2006,md2008,takato2008,
vok2009,cellino2010,willman2010,nov2010ch,ziffer2011}. Still, the
search for new young families is very important in many
respects. For instance, there is a lack of young and dynamically
stable groups belonging to the taxonomic $C$-class, as was noted by
\citet{nov2012lorre}, who found the first such example.

Another reason why young asteroid families are
important is their likely relation with a new class of asteroids
identified in recent years, collectively known as \textit{active
asteroids} \citep{jewitt2012}. Active asteroids are objects which move
along typical asteroid orbits, but exhibit observable
comet-like activity, i.e., mass loss, due to one or more of
different physical mechanisms as discussed by \citet{jewitt2012}. The two
most plausible explanations for the activity observed for most active asteroids are the sublimation of water ice and
the impulsive ejection of material by an impact. The main belt
asteroids whose activity driver is most likely to be sublimation are
referred to as \textit{main-belt comets} \citep[MBCs;][]{hsieh2006}.
Objects displaying likely impact-driven activity are known as
\textit{impacted asteroids} or \textit{disrupted asteroids}.

The existence of active asteroids, and MBCs in particular, challenges
the traditional view that asteroids and comets are two distinct
populations, and supports asteroid-comet continuum hypotheses \citep[e.g.,][]{gounelle2008,briani2011}.
So far, only a little more than a dozen active asteroids have been
discovered, but their number is constantly increasing with ongoing
survey work (e.g., by the Catalina Sky Survey, Pan-STARRS, and others)
and the improvement of the telescopes, detectors, and automated
comet-detection algorithms used in such surveys.

The assumption that MBC activity is driven by volatile sublimation
implies that volatile compounds, i.e., ices, must be present on or
immediately beneath the surfaces of these objects. It is difficult,
however, to explain the survival of ices or
other volatiles on (or close to) the surfaces of objects orbiting at
the heliocentric distances of the main asteroid belt over Gyr
time-scales \citep{hsieh2009,capria2012}. In fact, sublimation is
expected to deplete the volatile content of the external layers of main belt
objects over much shorter time scales. Hence, it has been suggested
that MBCs could be preferentially found among young asteroid families, 
since the recently-formed members of these young families could
still retain significant reservoirs of volatile material immediately below their surfaces
which were previously deeply buried in the interior of the
original parent bodies \citep{nes2008,hsieh2009,nov2012a}.

So far, links between MBCs and young families have been shown
in only two cases. 133P/Elst-Pizarro belongs to the
young Beagle family, which is estimated to be less than 10~Myr old
\citep{nes2008}. The second example is that of P/2006 VW$_{139}$,
which is a member of a small cluster of objects estimated
to be just 7.5 Myr old \citep{nov2012a}. If more cases of MBCs belonging
to young families can be found, it would lend strong support to
the hypothesis that these families could preferentially contain more MBCs
than the general asteroid population, which could in turn lead to
more efficient searches for even more MBCs and also to greater
insights into the physical conditions that give rise to MBC activity.
Thus, each time a new MBC is discovered, it is extremely important to check
whether or not an associated young asteroid family can be found.

Another reason why one could expect a cometary activity to be shared
by different members of a very young family is that a statistical
analysis has shown the occurrence of a strong enhancement in the
rate of mutual, low-energy collisions among the members of
newly-formed families \citep{dell'oro2002}. Although the period during
which the intra-member collision rate is enhanced over the background
collision rate is found to last only a
relatively short time, and is expected to have only a minimal effect on
the long-term collisional evolution of the family, this effect could nonetheless
have consequences on the cratering record on the surfaces of
family members, and could potentially enhance the likelihood of comet-like
activity arising on these objects.

Active asteroid P/2012~F5 (Gibbs) (hereafter P/2012~F5) was
discovered last year in Mt. Lemmon Survey data
\citep{gibbs2012}. To date, it has been the subject of two published studies
\citep{stevenson2012,moreno2012}, both suggesting it is a disrupted
asteroid, rather than a MBC.  Before the origin of the activity of P/2012~F5
had been conclusively determined, however, we had already begun the search for
a dynamical family associated with the object for the reasons described above.
We describe the results of that search in this paper.

The paper is organized as follows. First, in Section~\ref{s:hcm}, we
compute both the osculating and proper orbital elements of P/2012~F5.
We then employ the hierarchical clustering method to search
for a family around P/2012~F5, successfully identifying an associated
young asteroid cluster that we have named the Gibbs cluster. In
Section~\ref{s:age}, we determine the age of the Gibbs cluster,
and in Section~\ref{s:phy} we analyze
some of its physical properties. In
Section~\ref{s:obs}, we present the results of an observational search
for new MBCs in the region occupied by the cluster, and finally, in
Section~\ref{s:conclusions}, we discuss our results and conclusions.

\section{Search for a dynamical family associated with P/2012~F5 Gibbs}
\label{s:hcm}

\subsection{Determination of orbital elements}

To study the dynamical environment of P/2012~F5, we need reasonably
good orbital elements for the object. This includes both osculating
and proper elements. However, shortly after its
discovery, the orbit of P/2012~F5 was still characterized by relatively large
uncertainties. Thus, we made an effort to improve this situation as much as
possible.

For the purpose of orbit determination, we used three sets of
astrometric observations collected over a period of $\sim$3.6 years
from 2009 September 17 to 2013 May 12. The largest portion of the
dataset consists of $125$ observations obtained by various observing
stations during the discovery apparition in 2012. In addition, $7$
recovery observations were obtained by the authors in 2013 using
the 3.6~m Canada-France-Hawaii Telescope and the University of
Hawaii 2.2~m telescope on Mauna Kea. Finally, $17$ precovery
observations of this object were also found in Pan-STARRS1 survey
data, and were submitted to the Minor Planet Center by the authors,
adding an additional 2.4 years to the total observed arc for this object.

This total sample of $149$ observations has been used in this work
to carry out a refined determination of P/2012~F5's orbit. The full dataset
was first fit to a purely gravitational orbit by weighting every
observation according to the average historical performances of the
observational station that obtained it. Gravitational perturbations
for all of the major planets and the three most massive main belt asteroids
were included in the computation. Astrometric residuals for each
astrometric position were then computed, and observations showing an
offset in excess of $2''$ were removed from the sample used to
obtain the solution. A new orbit was then computed, and this
iterative process was repeated until a stable solution was achieved.
The final solution was found after rejecting $37$ out of the $149$
observations, where all of the rejected observations were obtained in
2012 when the object was active. The anomalous abundance of large outliers
during this period is
likely due to the peculiar morphology of the object
during its active phase, since the long tail structure and the lack of
a clearly defined central condensation made it difficult to locate
the object's photocenter (especially for small-aperture telescopes), in
turn causing a significant number of inaccurate positions to be
reported to the Minor Planet Center. In fact, among the rejected
positions, some show astrometric residuals in excess of $10$ arcsec,
mostly in the tail-ward direction. The resulting osculating orbit
(Table~\ref{t:ele}) includes $112$ positions, spanning an arc of
1333 days, and has an RMS of about $0\farcs656$. The addition of
non-gravitational terms following the usual A1/A2 formalism does not
improve the orbital fit, and no significant detection of
non-gravitational accelerations can therefore be extracted from this
dataset.

\begin{table*}
\footnotesize
 \centering
  \caption{The osculating orbit parameters and their corresponding formal errors at epoch JDT 2456400.5 (2013 Apr 18.0 TT)
  for active asteroid P/2012~F5.} \label{t:ele}
  \begin{tabular}{lcrrc}
\hline
Orbital element & Symbol & Value & Error & Units\\
\hline
Semi-major axis         & $a$      & 3.0050440      & 9.78e-7 & AU \\
Eccentricity            & $e$      & 0.0417036      & 2.22e-7 &  - \\
Inclination             & $i$      & 9.73869        & 0.000026& deg \\
Argument of perihelion  & $\omega$ & 177.82221      & 0.0007  & deg \\
Longitude of node       & $\Omega$ & 216.85955      & 0.00012 & deg \\
Mean anomaly            & $M$      & 210.98151      & 0.0008  & deg \\
Perihelion distance     & $q$      & 2.87972262     & 1.21e-6 & AU \\
Aphelion distance       & $Q$      & 3.13036534     & 1.15e-6 & AU \\
Perihelion passage      & $t_{p}$  & 2457188.112002 & 0.00472 & JD \\
\hline
\end{tabular}
\end{table*}

Having obtained good osculating elements we can then proceed to the
determination of proper orbital elements, which we did by applying
the methodology developed by \citet{MilKne90,MilKne94}.
We compute P/2012~F5's proper semi-major axis ($a_{p}$),
eccentricity ($e_{p}$), and inclination ($i_{p}$) and
list their values in Table~\ref{t:list}.

\subsection{Application of the Hierarchical Clustering Method}

The next step in our study was to search for the presence
of a dynamical asteroid family around P/2012~F5. Families
form dense clusters in the three-dimensional space of proper
semi-major axis ($a_{p}$), proper eccentricity ($e_{p}$), and
proper inclination ($i_{p}$), and can be identified by analyses
of the distribution of objects in proper element space to search
for such clusterings. In our analysis,
we apply the hierarchical clustering method (HCM) to perform this
analysis and adopt the `standard' metric, $d_c$, proposed by
\citep{zappala1990,zappala1994}.

The HCM identifies groupings of objects having mutual separations
below a threshold `distance' ($d_{c}$), which, adopting standard
conventions, has units of m~s$^{-1}$. We apply the HCM to a catalog
of analytically-determined proper elements \citep{MilKne90,MilKne94}
available at the \textit{AstDyS} web repository as of November 2012
(http://hamilton.dm.unipi.it/astdys/). Analytical proper elements
are reasonably accurate for objects with low to moderate orbital eccentricity
and inclination. We use them because they are available for both
numbered and multi-opposition asteroids. The proper elements of
P/2012~F5 that we obtained here are also added to the catalog.

\begin{figure}
 \centering
\includegraphics[angle=-90,scale=.4]{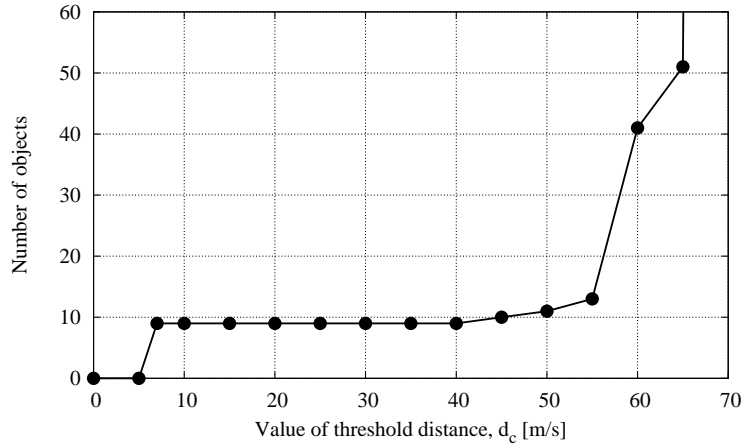}
\caption{Number of asteroids associated with P/2012~F5 as function of
cut-off distance (in velocity space), expressed in m~s$^{-1}$. The
dominant feature is the existence of a small group consisting of
nine members. These objects are very tightly packed in proper orbital
element space.}\label{f:nfv}
\end{figure}

We carry out our HCM analysis by testing a range of cutoff distances
from $5$ to $70$ m~s$^{-1}$ and noting the number of asteroids that
the analysis links to P/2012~F5 at those separations.
At the beginning of our search, we
change $d_{c}$ in discrete steps of 1 m~s$^{-1}$, but after
identifying the family at 7 m~s$^{-1}$, we switch to steps of 5
m~s$^{-1}$. Our results are shown in Figure~\ref{f:nfv}.  We
find that a cluster of asteroids around P/2012~F5 does indeed exist,
and hereafter will refer to it as the Gibbs
cluster\footnote{The usual practice is to name asteroid families
after their lowest numbered member. However,
for groups that are known to contain an active asteroid and that
are discovered as the result of a search around that active
asteroid, we have decided to
name them after the member known to be active.}. This cluster is extremely
compact and is clearly separated from background objects in
proper element space.  Given how compact the core of the cluster is,
we do not believe that the asteroids associated with the
cluster for $d_{c} >$ 40 m~s$^{-1}$ are real members.

\begin{table*}
\begin{minipage}{144mm}
\footnotesize
 \centering
  \caption{List of asteroids belonging to the Gibbs cluster.} \label{t:list}
  \begin{tabular}{lccccrrc}
\hline
Asteroid\footnote{Asteroid number or provisional designation}
& $a_{p}$\footnote{Proper semi-major axis in AU}
& $e_{p}$\footnote{Proper eccentricity}
& $sin(i_{p})$\footnote{Sine of proper inclination}
& $H$\footnote{Absolute magnitude}
& $D_{1}$\footnote{Diameter in km, when an albedo of $p_v$ = 0.05 is assumed}
& $D_{2}$\footnote{Diameter in km, when an albedo of $p_v$ = 0.2 is assumed}
& $T_{lyap}$\footnote{Lyapunov time in Myr} \\
\hline
20674            & 3.00423 & 0.02324 & 0.17974 & 12.6 & 17.9 &  9.0 & 0.65 \\
140429           & 3.00381 & 0.02315 & 0.17972 & 15.0 &  5.9 &  3.0 & 3.33 \\
177075           & 3.00509 & 0.02289 & 0.17973 & 15.6 &  4.5 &  2.3 & 0.63 \\
249738           & 3.00481 & 0.02294 & 0.17971 & 15.7 &  4.3 &  2.2 & 0.68 \\
257134           & 3.00514 & 0.02300 & 0.17970 & 15.8 &  4.1 &  2.1 & 0.67 \\
321490           & 3.00514 & 0.02299 & 0.17971 & 15.8 &  4.1 &  2.1 & 0.65 \\
2007~RT$_{138}$  & 3.00484 & 0.02310 & 0.17969 & 15.7 &  4.3 &  2.2 & 0.65 \\   
2002~TF$_{325}$  & 3.00503 & 0.02288 & 0.17968 & 17.1 &  2.3 &  1.1 & 0.67 \\
P/2012 F5        & 3.00386 & 0.02274 & 0.17972 & 17.4 &  2.0 &  1.0 & 0.83 \\
\hline
\end{tabular}
\end{minipage}
\end{table*}

\begin{figure}[ht!]
 \centering
\includegraphics[angle=-90,scale=.29]{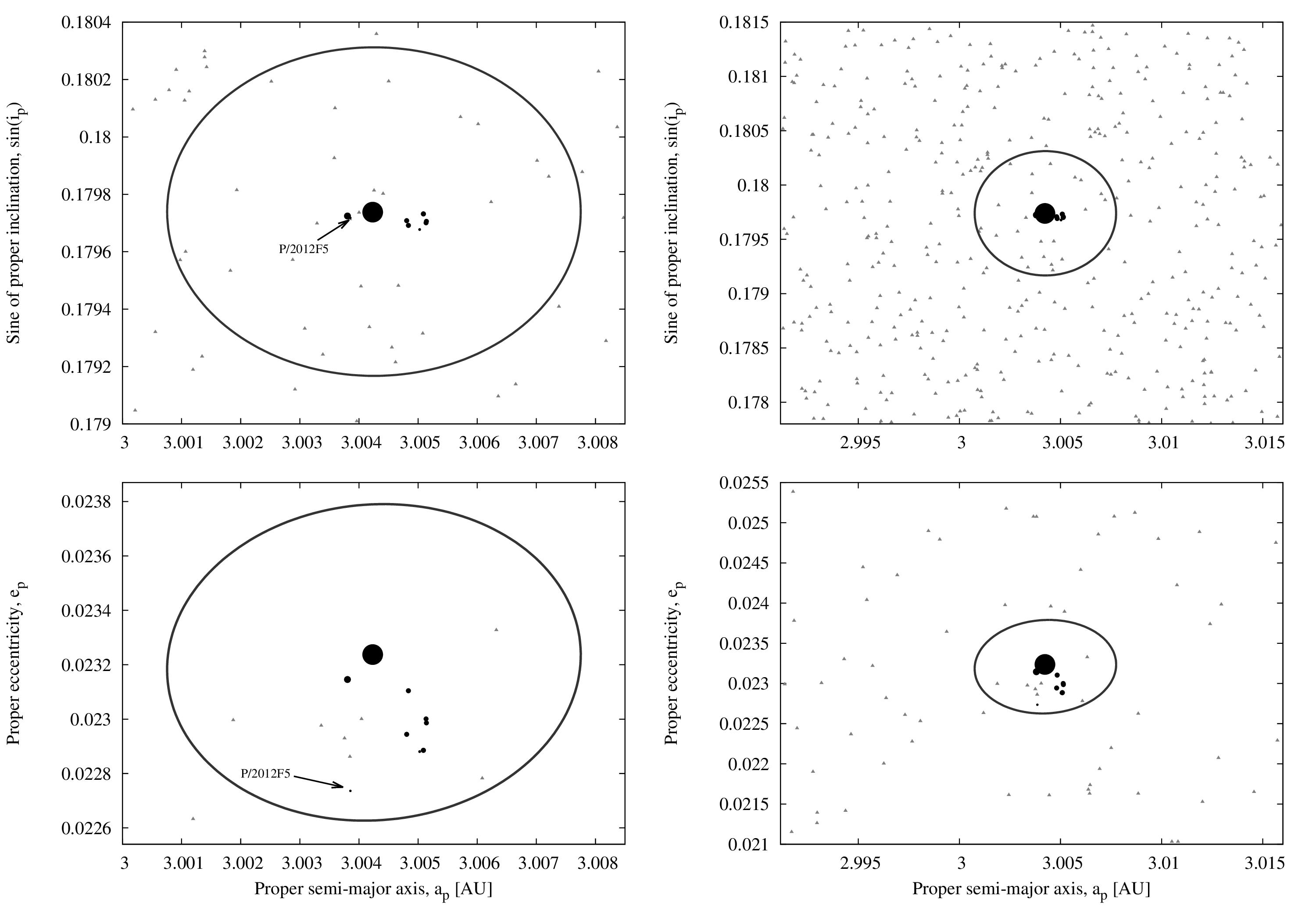}
\caption{The region of the main asteroid belt in which the Gibbs cluster is
located. The plots represent the space of proper orbital elements,
in two different planes (top/bottom) and scales (left/right). The
members of the cluster are shown as black filled circles, and the
size of their symbols is proportional to the corresponding diameter.
Nearby background asteroids are shown as gray filled triangles.
For the meaning of the elliptical curves, see the text.
}\label{f:aei_all}
\end{figure}

The structure of the cluster in the space of proper orbital elements
is shown in Figure~\ref{f:aei_all}. In this figure, all asteroids
located in the region of the Gibbs cluster are shown in two planes
(semi-major axis \emph{vs.} eccentricity and semi-major axis
\emph{vs.} sine of inclination), and by using two different scales.
In the plots, the superimposed ellipses represent equivelocity
curves, computed according to the Gauss equations \citep{morby1995}.
These ellipses are obtained assuming a velocity change $\Delta v =
10$ m~s$^{-1}$, argument of perihelion $\omega = 90^{o}$, and true
anomaly $f = 90^{o}$. The ellipses are shown as an illustration of
the limiting distance between the parent body and the other fragments in
the isotropic ejection field. However, the ejection field of
the Gibbs cluster is clearly asymmetric, a property that is usually
interpreted as indicating that a family is the
result of a cratering event \citep{vok2011,nov2012lorre}.

\subsection{Statistical significance of the cluster}

An important additional step in any family identification analysis is
evaluation of the statistical significance of any identified groups in
order to avoid confusing true families (i.e., clusters
of asteroids sharing a mutual collisional origin) with
groupings which are simply statistical flukes.
To evaluate the significance of the Gibbs cluster, we first note that
its members are extremely tightly packed in proper orbital element space.
We further note that the density of asteroids in the immediate vicinity of the
Gibbs cluster is relatively low. This can be seen by looking at the
four plots shown in Figure~\ref{f:aei_all}. There are a few
background asteroids located inside the equivelocity ellipses in
one plot or another, but these are not the same objects in both planes.
As such, none of these objects are actually located within the proper
element space region defined by the cluster.

Relatively close to the Gibbs cluster, there is the very large Eos
family. This family is among the largest and oldest groups in the
main belt \citep{vok2006}. However, the eccentricities of asteroids
belonging to the Gibbs cluster are substantially lower than those of
Eos family members, even when a possible external halo of Eos
family members is considered \citep{broz2013}.  We therefore consider
the Gibbs cluster to be clearly separated from the Eos family, and likely
completely unassociated with it. Moreover, the little spectroscopic data
available for the Gibbs cluster also seem to rule out any relation
with the Eos family (discussed below).

\citet{karin2002} furthermore showed that, even within the borders of the large and dense Koronis
family, using a very low critical distance threshold for family
identification of $d_{c} = 10$~m~s$^{-1}$, clusters of only up to 5
members could be found by chance.  This result suggests that a concentration
of asteroids as tight and dense as the Gibbs cluster is not easily achievable,
even within very densely populated volumes of proper element space, including those
occupied by extremely large asteroid families like Eos, Themis and
Koronis, further suggesting that the Gibbs cluster is a true asteroid family,
and not a statistical fluke, and that its members share a
common collisional origin.

To make a more quantitative assessment of the significance of the Gibbs
cluster, we also perform the following test.
First, in the space of proper orbital elements, we generate $1000$
different synthetic main asteroid belts, each one including
336~555 fictitious objects drawn from a quasi-random distribution
(QRD) fitting the distributions of $a_{p}$, $e_{p}$, and
$\sin(I_{p})$ exhibited by the known asteroids in the
real main asteroid belt. By doing this, we are able to experiment with
different random populations while still taking into account the structure of
the real asteroid belt. The complete procedure to obtain the QRD is described 
in \citet{nov2011}.
We then apply the HCM to each of our $1000$ synthetic main
belts. Using the cut-off distance of 7~m~s$^{-1}$ (the level at
which the Gibbs cluster is detected), we fail to find any group
with at least 9 members. We therefore conclude that the statistical
significance of the Gibbs cluster is $>99.9$\%.

Despite the results presented above, one should keep in mind that the high statistical significance of
the cluster itself dose not imply that there are no interlopers. A priori we cannot exclude a possibility
that any single member is an interloper.

\section{Age of the Gibbs cluster}
\label{s:age}

The most appropriate and accurate way to determine the age of a young
family is the so-called backward integration method (BIM;
\citet{karin2002}) The strategy behind the BIM relies on the fact that immediately
after the disruption of a parent body, the orbits of the fragments
are nearly identical (being determined by the ejection velocities
through the Gauss equations), but then tend to diverge as a
function of time due to planetary perturbations and
non-gravitational effects. Consequently, two secular angles that
determine the orientation of an orbit in space, namely the longitude
of the ascending node ($\Omega$) and the argument of perihelion
($\omega$), for different objects evolve with different, but nearly
constant, speeds. After some time, this effect tends to spread out
the distributions of $\Omega$ and $\omega$
of the family members uniformly over $360^{\circ}$. Therefore, the age of a young asteroid
family can be determined by numerically integrating the orbits of
its members backwards in time until the orbital orientation angles
cluster around single values. Of course, this can be reliably done
only when a family is sufficiently young that the
dynamical evolution of its members, following fragmentation of the parent body, 
has not yet completely erased information about the primordial orbits.

This method, either in its original form or with some variations, has
been used many times in the last decade to estimate the age of young
families. For example, the BIM has been used to determine the ages of the Karin
cluster \citep{karin2002}, Veritas family
\citep{nes2003}, Datura cluster \citep{datura2006}, Theobalda family
\citep{nov2010theo}, and Lorre cluster
\citep{nov2012lorre}. Here, we obtain the
age of Gibbs cluster using two approaches based on the BIM. First,
we determine the cluster's age by numerically integrating the orbits of the
nominal cluster members, as originally proposed by
\citet{karin2002}. Second, we refine our estimate using orbital and
Yarkovsky clones of the real cluster members, in a way similar to
that proposed by \citet{vok2011}.

\subsection{Orbital evolution of the cluster members}

In our first application of the BIM, only the orbits of real family
members are used to estimate the family age. In order to apply the
BIM, two conditions must be fulfilled: (1) the family must be young
(up to about 10~Myr); and (2) the family members must be dynamically stable.
The first condition is nearly certainly satisfied based on the very tight
packing of the Gibbs cluster members in the proper orbital space as
discussed in Section~\ref{s:hcm}. The fulfillment of the second
condition is verified by calculating Lyapunov times ($T_{lyap}$) for
all objects belonging to the cluster. In practical terms, for the
purpose of this study, objects are considered stable if $T_{lyap} >
10^{5}$~yr. This condition is satisfied for all members of the cluster
(see Table~\ref{t:list}).

\begin{figure}[ht!]
\centering
\includegraphics[angle=-90,scale=.35]{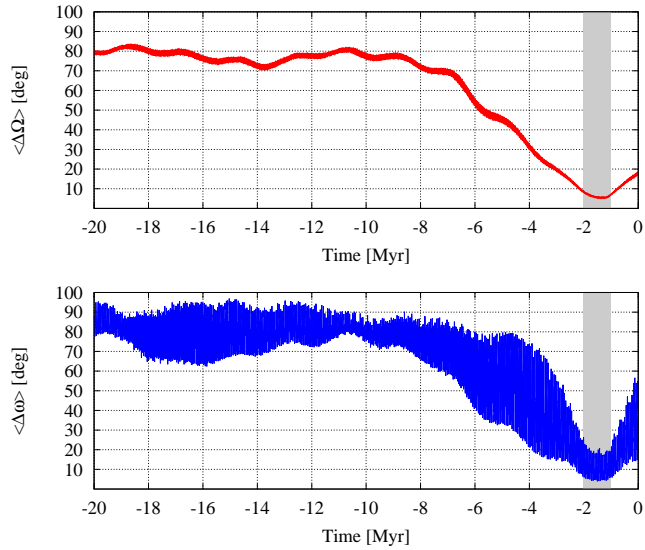}
\caption{The average of differences in the mean longitudes of the
ascending nodes $\Omega$ (top), and arguments of perihelion $\omega$
(bottom), for the $9$ nominal members of the Gibbs cluster. These
results are obtained in a purely gravitational model. The most
important feature (clearly visible in both plots) is a deep
clustering of both angles occurring about 1.5~Myr in the
past.}\label{f:age_nominal}
\end{figure}

\begin{figure}[ht!]
\centering
\includegraphics[angle=-90,scale=.35]{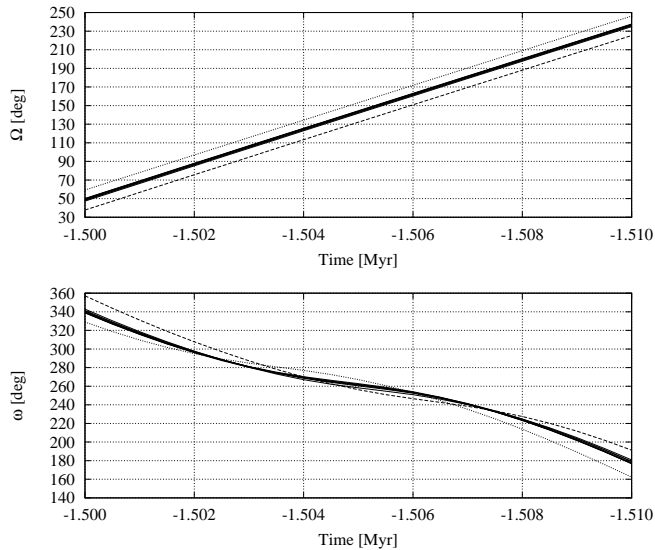}
\caption{The plot shows past orbital evolutions of nodal longitudes, $\Omega$ (top) 
and arguments of perihelion, $\omega$ (bottom) at about 1.5 Myr ago for all 9 members
of the Gibbs cluster. At this time both angles of all 9 asteroids were nearly the same, suggesting that the
cluster was created by a recent catastrophic collision.}\label{f:orbits_evolution}
\end{figure}

The evolution of the average of the mean differences in the two
secular angles, derived from the numerical integration of the orbits
of the cluster members, is shown in Figure~\ref{f:age_nominal}. The
results clearly show a tight clustering (within $7$ degrees) of both
angles at $\sim1.5$~Myr in the past. This clustering very likely
corresponds to the time of family formation. Such a conclusion is additionally
supported by the past evolution of individual orbits of all 9 members of the cluster.
As it is shown in Fig.~\ref{f:orbits_evolution}, both secular angles,
the nodal longitudes and arguments of perihelion, were very close at this
time. Furthermore, the latter result also indicates that all 9 asteroids are likely real members
of the Gibbs cluster.

\subsection{Orbital and Yarkovsky clones}

To further refine our BIM determination of the age of the cluster, we
use a methodology first proposed by \citet{nes2006} and
used in the past to estimate the ages of some young
clusters \citep[e.g.,][]{vok2011,nov2012lorre}. Thus, we refer
the reader to these papers for additional information about the
method. In principle, the BIM is affected by two major sources of
error. These are due on one hand to unavoidable uncertainties in the
orbital elements of known family members, and on the other hand to a
well-known secular evolution of the semi-major axis caused by the
Yarkovsky thermal force \citep{bottke2006}. The latter depends, in
turn, upon the thermal properties of the objects' surfaces and on
the value of the obliquity angle. As a consequence
of the Yarkovsky mechanism, the semi-major axis can either increase
or decrease with time. In order to account for the above effects, we
extend our analysis by considering a large sample of
synthetic clones.

For this analysis, we generate a set of statistically equivalent orbital
and Yarkovsky (hereafter 'yarko') clones. Specifically, for each
nominal member of the Gibbs cluster, we create $10$ orbital clones,
and for each of the orbital clones, we generate $10$ different yarko
clones corresponding to different possible drift rates of the
orbital semi-major axis. Orbital clones are generated using
$3\sigma$ uncertainties of each cluster member's osculating orbital
elements, assuming Gaussian distributions.\footnote{A better way to produce 
orbital clones would be to use random distribution based on the full
correlation matrix. The approach we used here makes
clones somewhat more dispersed, resulting in slightly larger 
uncertainty of the age than necessary. Still, we used this method due to its simplicity.}
Yarko clones are distributed randomly over the interval $\pm (da/dt)_{max}$, where
$(da/dt)_{max}$ is the maximum expected value of the semi-major axis
drift speed caused by the Yarkovsky effect. This random $da/dt$
distribution corresponds to an isotropic distribution of spin
axes. At the location of the
Gibbs cluster, for a body of $D=1$~km in diameter, we use a value
of $(da/dt)_{max} = 4 \times 10^{-4}$~AU/Myr, which scales as $1/D$.
This drift limit was determined assuming thermal and physical parameters
appropriate for C-type asteroids \citep[see e.g.][]{broz2008}.
Note also that we only take into account the diurnal component of the
Yarkovsky effect, because the seasonal variant is negligible
for the objects of these sizes \citep{bottke2006}.
In this way, we assign a total of $100$ statistically equivalent clones
to each real member of the cluster.

We then numerically integrate the orbits of all clones backward
in time for $2$~Myr using the \textit{ORBIT9} software package
\citep{orbit9}. The adopted dynamical model includes four major
planets, from Jupiter to Neptune, as perturbing bodies, and accounts
for the Yarkovsky effect.

The age of the cluster is defined as the minimum of the function
\begin{equation}
\Delta V = n a \sqrt{(sin(i) \Delta \Omega)^{2} + 0.5(e \Delta \varpi)^{2}}
\label{eq:fun}
\end{equation}
where $na \approx$ 17.2 km~s$^{-1}$ is the mean orbital speed of the
asteroids in the Gibbs cluster, and $\Delta \Omega$ and $\Delta
\varpi$ are the dispersions of the longitude of node and the
longitude of perihelion, respectively \citep{vok2011}. We then obtain
the final age of the cluster by
performing $10^{6}$ trials of this procedure, randomly selecting one clone of each
member, and determining the minimum of the function defined above
for all of the clone combinations.

\begin{figure}[ht!]
 \centering
\includegraphics[angle=-90,scale=.4]{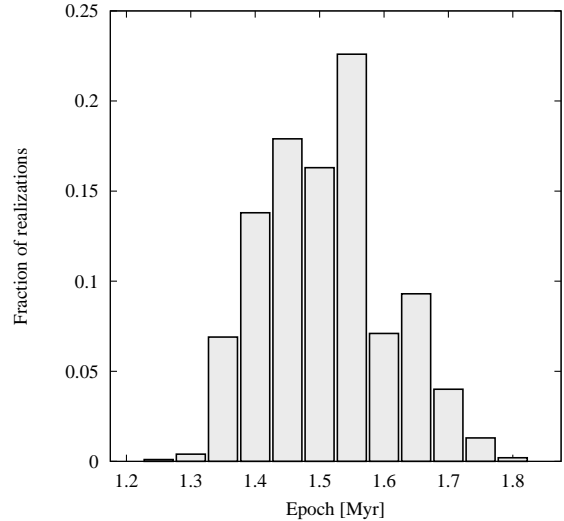}
\caption{Histogram of possible ages of the Gibbs cluster, produced
using $10^{6}$ different combinations of orbital and Yarkovsky
clones. }
\label{f:age}
\end{figure}

The histogram of the ages we obtain using this method is shown in Figure~\ref{f:age}. 
We find the age of the Gibbs cluster to be $1.5\pm0.1$~Myr.

\section{Physical properties of the Gibbs cluster}
\label{s:phy}

Unfortunately, very little physical and spectral information about the
members of the Gibbs cluster exists. The only data at our disposal come
from SDSS spectrophotometric observations and corresponding
taxonomic classification. Specifically, Gibbs cluster members
(140429) 2001TQ$_{96}$ and (177075) 2003FR$_{36}$ are both
classified as $Q$-class objects, although with probabilities of only
32\% and 13\% respectively \citep[see][for more details on the
classification]{carvano2010}. If these classifications are correct, we
might expect all Gibbs cluster members to be $Q$-type asteroids since
the members of an asteroid family tend to
share similar spectral properties \citep[e.g.][]{cellino2002}.

This would be a very interesting result because $Q$-type asteroids are
spectroscopically more similar to ordinary chondrite meteorites than
any other asteroid class \citep{bus2002}, and it has been suggested that
$Q$-class asteroids, which are most common among near-Earth objects
(NEOs), have young surfaces. As a consequence, space
weathering \citep[SW;][]{gaffey2010,marchi2012} has presumably not had 
sufficient time to transform the surfaces of $Q$-class
asteroids into those typical of classical and more common $S$-type
asteroids. The timescale for an asteroid to transition from $Q$-type to $S$-type is
still not well understood. \citet{cellino2010} did not find
differences in polarimetric properties of equal-sized members of the
Koronis and Karin families, implying that SW acts on timescales
shorter than the age of the Karin family, or about $6$ Myr.
\citet{vernazza2009} claimed that SW
rapidly reddens asteroid surfaces in less than 1~Myr. However, this may be
inconsistent with the observed fraction of $Q$-type
asteroids among NEOs. To reconcile this problem, \citet{binzel2010}
proposed that during close encounters between NEOs and the
Earth-Moon system, tidal forces could cause surface shaking that
rejuvenates the surface regolith of these objects, thus, 
returning them from $S$-types back to $Q$-types. On the other hand, recent work by
\citet{nes2010} suggests a timescale longer than 1~Myr for SW
to affect NEO spectra, so there may be no conflict to resolve after all.

Interestingly, the first examples of main belt $Q$-type asteroids
were found just recently among the members of very young
asteroid families \citep{md2008}. In this respect, our findings
for the Gibbs cluster could be interpreted as an indication
that, at these heliocentric distances, SW mechanisms require longer
than $1.5$ Myrs (our derived age of the family) to change the
spectrophotometric properties of bodies having an overall
composition similar to that of ordinary chondrites. However, taxonomic
classifications based on multi-band photometry covering only a few
color channels are obviously not as precise as classifications derived from
full reflectance spectra, and so therefore caution is required to avoid
over-interpreting the scarce data at our disposal.

We note that \citet[][]{rivkin2011} analyzed some
members of the Koronis family (about 1-2 billion years old) and found that $Q$-type 
asteroids are also present in the main-belt among asteroids smaller than about 4~km
\citep[see also][]{thomas2011,thomas2012}.
These results are not necessarily at odds with our
conclusions about possibly young surfaces among Gibbs
cluster members, but they do perhaps indicate that more work is needed to better
understand the nature of $Q$-type asteroids.

Other physical parameters can be derived for the members of the
Gibbs cluster based on their absolute magnitude values taken from
catalogs, and making assumptions about albedo and composition.
Given the uncertainties on the few available spectrophotometric
data, we decided to assume two different values for the geometric
albedo $p_{v}$ and the density $\rho$, assuming two possible physical
situations. As such, we consider the case of primitive
$C$-type objects having low albedo and density ($p_{v}=0.05$ and
$\rho=1.3$ g~cm$^{-3}$) and the case of $S$-type
asteroids with higher albedos and densities ($p_{v}=0.2$ and $\rho=2.5$
g~cm$^{-3}$) \citep{carry2012}.

First, for both sets of physical parameters, we compute the diameters of all
cluster members using the absolute magnitudes provided by the
\textit{AstDyS} website. These results are also given in Table~\ref{t:list}.
It should be noted that the largest member, asteroid (20674)
1999~VT$_{1}$, is significantly larger than all other members,
about 3 times larger than the second largest
member, asteroid (140429) 2001~TQ$_{96}$.  This situation is unlikely
to be the consequence of observational incompleteness because
all asteroids with $H<15$ at heliocentric distances of $\sim$3~AU are
believed to have been discovered \citep{gladman2009}.

Next, we estimate a lower limit for the diameter of the parent body
$D_{PB}$ (assuming a spherical shape) by summing-up the volumes of
all known members. We find that $D_{PB}\geq18.3$~km or $D_{PB}\geq9.1$~km,
depending on the albedos assumed for cluster members. Corresponding
escape velocities are $\sim7.8$~m~s$^{-1}$ or $\sim5.4$~m~s$^{-1}$, respectively for
the two cases, taking into account the assumed density values
mentioned above. We also note that if we follow the arguments
developed by \citet{tangaetal99}, the parent body's diameter could
not have been less than the sum of the sizes of the two largest
members. The resulting parent body size turns out therefore to be on
about $24$~km (assuming $p_V=0.05$) or about $12$~km (assuming $p_V=0.2$).

As we have already noted, asteroid 20674 is by far the largest member
of the Gibbs cluster. The mass ratio between the largest fragment and the
parent body is $M_{LF}/M_{PB}\approx0.9$. Based on these findings,
the collision producing the Gibbs cluster should be considered to
be between a catastrophic disruption and a cratering
event. Additional information, such as the discovery of more cluster members
or better observational data for the cluster members,
is certainly needed to better constrain the nature of the initial family-forming
fragmentation event.

\section{The observations}
\label{s:obs}

\subsection{Search for new main-belt comets}

Young asteroid families located in the outer regions of the asteroid
main belt are thought to be the best candidates to look for new
main-belt comets \citep{hsieh2009}. For these reasons, we have
carried out observations of $5$ members of the Gibbs cluster and of
$2$ additional nearby background objects.

From the observational point of view, the major difficulty in
identifying new MBCs is that of being able to detect their elusive
cometary-like activity that is both weak and transient. Several techniques to attack this problem
have been used so far \citep[see][and references
therein]{hsieh2009,sonnett2011,Waszczak2013}.

The approach that we followed here includes optical imaging and
adopts two different search methods. The first method is based on
the so-called Stellarity Index, derived from SExtractor
\citep{bertin1996}. This is designed to discriminate between the
images of point-like and extended sources. The second method
consists in comparing surface brightness profiles of both the target
and a nearby reference star \citep{hsieh2005}. A possible excess in
an asteroid's profile, would be diagnostic of the presence of a
comet-like coma.

We obtained $R$-band imaging of seven targets using the
Imager/Low Resolution Spectrograph Do.Lo.Res of the 3.6~m Telescopio
Nazionale Galileo (TNG) located at the Observatorio del Roque de los
Muchachos (ORM) at La Palma, Canary Islands. Do.Lo.Res (Device
Optimized for the Low Resolution) is a focal reducer instrument
installed at the Nasmyth B focus of the TNG. The detector is a 2048
$\times$ 2048 E2V 4240 thinned back-illuminated, deep-depleted,
Astro-Broadband coated CCD with a pixel size of 13.5~$\mu$. The
plate scale is $0\farcs252$~pixel$^{-1}$, yielding a field of view of
about $8.6'\times8.6'$. Observations were carried out in
service/queue mode between August 2012 and January 2013.
Observational circumstances are listed in Table~\ref{t:obs1}.

\begin{table*}
\begin{minipage}{150mm}
\footnotesize
 \centering
  \caption{Observational circumstances of the 7 targets observed at the TNG. All observations were obtained using
$R$-Johnson filter.} \label{t:obs1}
  \begin{tabular}{lccccrcrrr}
\hline
UT date\footnote{Date of the observation}
& UT time\footnote{Universal time of the observation in hrs}
&Target\footnote{Asteroid (target) number or provisional designation}
& Ext.\footnote{Extinction in mag/airmass}
& DIMM\footnote{Differential Image Motion Monitor seeing in arcsec}
& t\footnote{Total integration time in sec}
& R\footnote{Heliocentric distance in AU}
& $\Delta$\footnote{Geocentric distance in AU}
& $\nu$\footnote{True anomaly in degrees}  \\
\hline
17/08/2012 & 2.5579 & 2002~TF$_{325}$ & 0.147 & 0.75 & 3600 & 3.001 & 1.999 & 271.4 \\
17/08/2012 & 3.8192 & 2007~RT$_{138}$ & 0.147 & 0.75 & 3600 & 2.989 & 2.019 & 278.8 \\
19/09/2012 & 1.5640 & 16290           & 0.129 & 0.85 & 3600 & 2.851 & 2.123 &  34.2 \\
19/09/2012 & 2.9158 & 82522           & 0.129 & 0.85 & 3600 & 2.889 & 2.206 &  37.9 \\
25/09/2012 & 4.3553 & 140429          & 0.171 & 0.90 & 3600 & 2.875 & 2.693 &  40.9 \\
12/01/2013 & 4.1661 & 177075          & 0.087 & 0.75 & 3600 & 3.025 & 2.283 & 107.8 \\
12/01/2013 & 5.5408 & 249738          & 0.087 & 0.75 & 3600 & 3.030 & 2.358 & 109.9 \\
\hline
\end{tabular}
\end{minipage}
\end{table*}

All the observing nights were photometric, with sub-arcsecond seeing.
Seeing data were available in real-time from the TNG DIMM and
extinction data in the SDSS $r$ band were available from the webpage
of the Carlsberg Meridian Circle telescope at the ORM. For all
targets, the same observation sequence was adopted of 12 exposures of
$300$~s each (total integration time of $3600$~s) while tracking each
asteroid with its proper differential motion. During each night, a
photometric standard star field \citep{landolt1992} was observed to
derive the average zero-point. We estimate the errors of the
photometric calibration of the fields to be $0.03$~mag or
less for all fields. Images were reduced following standard
procedures using IRAF routines. First, a master bias frame was
created for each night by averaging all bias frames obtained
that night. All images were then bias-corrected by subtracting the
corresponding master bias. A master flat-field frame was obtained by
averaging the bias-corrected flat-field images and normalizing to
the median intensity value. Images were then corrected for
pixel-to-pixel response variations dividing by the corresponding
flat-field frames. Given the non-negligible apparent differential motion of the
asteroids with respect to the background star field, a suitable non-zero
differential R.A./Dec. tracking rate was applied to the telescope
TCS for each target. This produces a field where only the asteroid
is a point-like source while all the other sources are trailed.
The final $3600$~s image of each asteroid was obtained by aligning,
registering and stacking each $300$~s image to the position of the
asteroid on the CCD corresponding as measured in the 6th image (i.e.
the middle of the acquisition sequence).

The next step is to check for possible signs of cometary activity
of the observed objects. Our first approach utilizes SExtractor
\citep{bertin1996}, a software package developed to detect, measure and
classify sources from astronomical images. Having been originally
designed to distinguish between stars and galaxies, it allows users
to discriminate between point-like and extended objects.

We use SExtractor to derive all the photometric/morphological
parameters such as flux, background level, $R$-band magnitude, FWHM,
ellipticity and Stellarity Index (SI). These results are shown in
Table~\ref{t:obs2}.

\begin{table*}
\begin{minipage}{150mm}
\footnotesize
 \centering
\caption{Photometric Data for our 7 target asteroids. The $R$-band
magnitude of each object is shown, as well as the 5$\sigma$ detection
limit magnitude for point-like and extended sources in each frame.
Stellarity Indices are reported to discriminate between point-like and
extended objects.} \label{t:obs2}
  \begin{tabular}{lccccc}
\hline
Target
& FWHM\footnote{FWHM in arcsec}
& $m_R$ \footnote{$R$-band apparent magnitude}
& Limit mag 1 \footnote{5$\sigma$ $R$-band detection limit magnitude for point-like sources}
& Limit mag 2 \footnote{5$\sigma$ $R$-band detection limit magnitude for extended sources in mag/arcsec$^{2}$}
& SI\footnote{Sextractor Stellarity Index} \\
\hline
2002~TF$_{325}$ & 0.92  & 21.13 & 24.61 & 25.14 & 0.98 \\
2007~RT$_{138}$ & 0.88  & 19.70 & 24.60 & 25.08 & 0.98 \\
16290     & 1.21  & 19.12 & 24.12 & 24.95 & 0.97 \\
82522     & 1.09  & 18.57 & 24.44 & 25.16 & 0.99 \\
140429    & 1.40  & 20.37 & 23.69 & 24.67 & 0.98 \\
177075    & 0.93  & 20.31 & 24.67 & 25.21 & 0.98 \\
249738    & 0.99  & 20.70 & 24.40 & 25.01 & 0.98 \\
\hline
\end{tabular}
\end{minipage}
\end{table*}

In particular, the latter parameter has been used to discriminate
between point-like and extended sources. This parameter is the
result of a supervised trained neural network to perform star-galaxy
classification. In theory, SExtractor considers objects with SI=0.0
to be galaxies and those with SI=1.0 to be stars. In practice,
objects are classified as stars by selecting SI $\geq$ 0.9. Since
the SI depends on the assumed FWHM of the stars in the image and the
$3600$~s exposures of our targets only have the target itself as the
sole non trailed source, we also acquired a short $20$~s exposure of
each field so that the differential tracking rate would not produce
any smearing of the stars and derived the average FWHM from that
image. We take into account any fluctuation of the seeing during the
$12 \times 300$~s sequence by checking the DIMM seeing. Despite a
differential tracking rate was applied to the TCS we found that $4$
out of $7$ targets do have a quite elongated PSF (ellipticity values
between 0.110 and 0.155). However, we attribute this elongation most
likely to the presence of aberrations in the telescope optics, since
no Shack-Hartman analysis was performed before observations. This is
why we adopted a flexible elliptical aperture \citep{kron1980}
instead of a simple circular aperture for photometry. Moreover, this
is the best choice when any object could have a intrinsic
diffused/elongated structure. All our targets are found to have SI
values of $\ge 0.97$. As a result, we conclude that none of the
asteroids observed in our program show any evidence of
cometary-like activity.

To confirm the above conclusions based on the Stellarity Index, we
also analyze all obtained images by comparing surface
brightness profiles of each asteroid and a corresponding reference
star. This technique is best suited for detecting coma that extends
radially in all directions from an object, or directed emission not
aligned with the direction of the object's apparent motion. It
cannot be used to detect emission oriented along the direction of an
object's apparent motion.

\begin{figure}[ht!]
\centering
\includegraphics[angle=0,scale=.4]{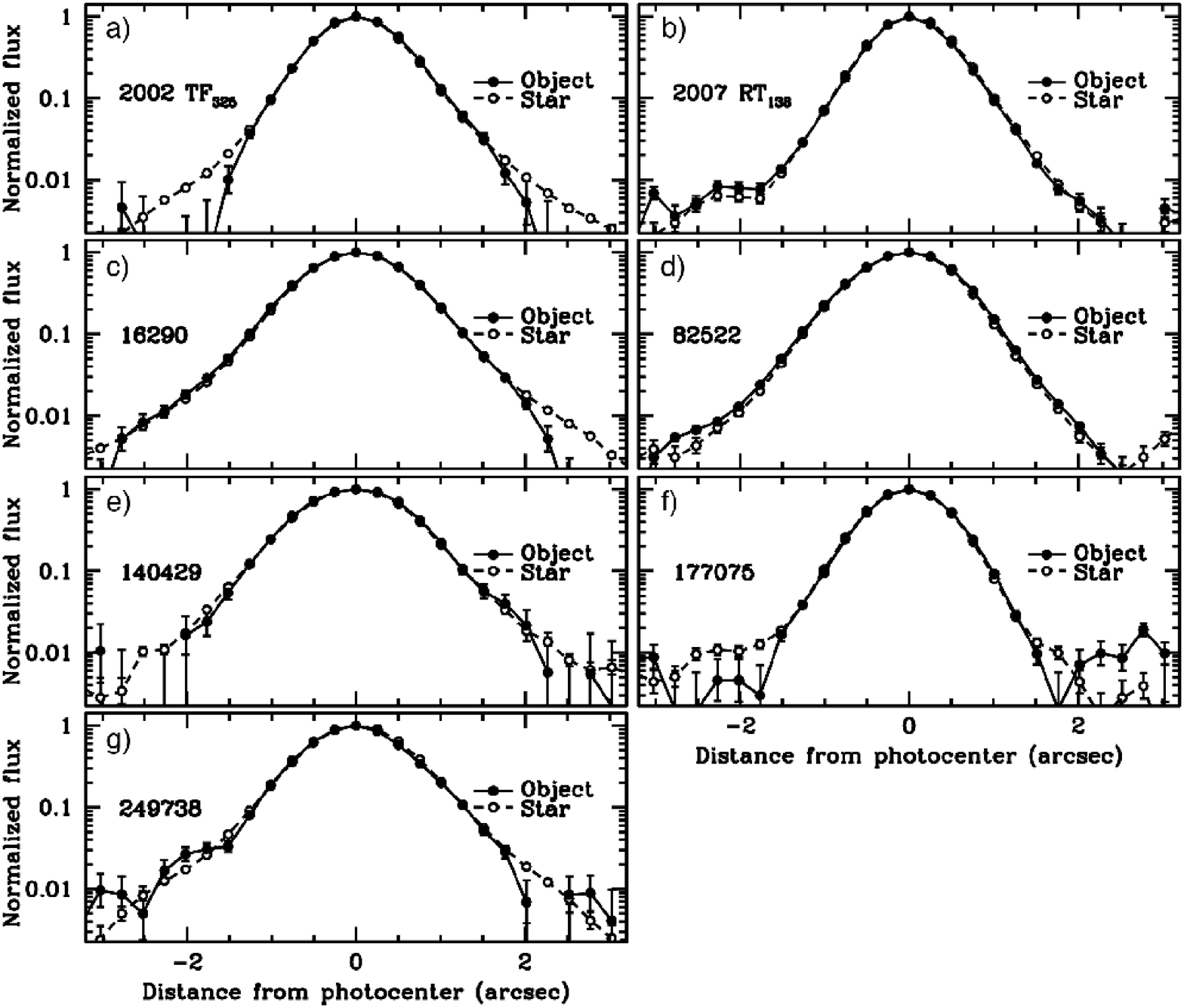}
\caption{Comparison of the surface brightness profiles of the composite images
of observed asteroids and corresponding reference stars. Surface brightness is normalized to unity
at each profile’s peak and is plotted on a logarithmic scale against angular distance
in the plane of the sky.}\label{f:obs_flux}
\end{figure}

We combine individual images of each object into a single high
signal-to-noise ratio composite image to search for any features
that would indicate comet-like activity. In each case, images are
shifted and aligned on corresponding object's photocenter using a
fifth-order polynomial interpolation and averaged. As was found
by \citet{hsieh2005}, this process produces less noisy profiles than
median combination. Additionally, all images are shifted and aligned on the
photocenter of a nearby reference field star. We can then obtain
one-dimensional surface brightness profiles by averaging over
horizontal rows over the entire widths of the object and reference
star, and by subtracting sky background sampled from either side of
the object or star. These profiles are then normalized to unity and
shown together in Figure~\ref{f:obs_flux} to search for
dissimilarities. Specifically, we looked for excess flux in each
asteroid's profile that would imply the presence of a coma. By
analyzing Figure~\ref{f:obs_flux}, we note that some scatter is present
in the wings of some of the asteroid profiles, but we attribute this to
low signal-to-noise ratios
far from the nucleus. Thus, we conclude that no coma is found, in
agreement with results we obtained using method based on the
SExtractor.

\subsection{Observations of P/2012~F5 Gibbs}

We also obtained $8$ $R$-band images totalling 4080~s of effective
exposure time on UT 2013 May 12, and 6 $R$-band images totalling
$1800$~s of effective exposure time on UT 2013 May 13 of P/2012~F5
using the University of Hawaii (UH) 2.2~m telescope on Mauna Kea.
Obtained under photometric conditions, these observations utilized a
Tektronix 2048$\times$2048 pixel CCD with an image scale of
$0\farcs219$~pixel$^{-1}$ and a Kron-Cousins $R$-band filter.
Standard bias subtraction and flat-field reduction were performed on
all images, where flat fields were constructed from dithered images
of the twilight sky.  Photometry of \citet{landolt1992} standard
stars was obtained by measuring net fluxes (over sky background)
within circular apertures, with background sampled from surrounding
circular annuli. Asteroid photometry was performed similarly, except
that to avoid the contaminating effects of any near-nucleus dust,
background sky statistics were measured manually in regions of blank
sky near, but not adjacent, to the object.  Several (5-10) field
stars in the object images were also measured and used to correct
for minor extinction variation during each night.

We use these observations to estimate the absolute magnitude of P/2012~F5,
finding $H_R\approx17.0\pm0.1$~mag and $H_V\approx17.4\pm0.1$~mag in $R$- and $V$-band 
respectively, assuming $G=0.15$ for both filters.
While the object appears point-source-like in
these images, very large dust grains ejected during P/2012 F5's original
outburst event in 2012 could have a dissipation rate from the nucleus that
is slow enough that they had not yet drifted beyond the seeing disk of the
nucleus at the time of our observations.  If this is the case, their
additional scattering surface area could have contributed to the total flux
observed from the nucleus, even though no observable evidence of residual
activity (either in the form of visible coma or a non-stellar PSF) was
present.  As such, we cannot absolutely rule out the presence of unresolved
large dust grains in the seeing disk of the nucleus.
Based on these absolute magnitude limits, we set an upper limit on the diameter of the 
nucleus of $\sim$2~km (see Table~\ref{t:list}).  

\section{Discussions and Conclusions}
\label{s:conclusions}

Our discovery of a young cluster associated with P/2012~F5
opens many different opportunities for future work. In this respect, three
characteristics of the cluster are of particular importance. First,
it is extremely compact in proper orbital elements, and its statistical significance is very high,
meaning that its members are very likely to be fragments originating from a
common parent body. Second, the Gibbs cluster is very young, being
only about 1.5~Myr old. Third, it is located in a dynamically cold
region of the main-belt, and thus its post-impact evolution is
bounded. The study of young and well preserved families like the Gibbs cluster
are essential for studies of impact physics, space weathering
effects, and dynamical evolution.

In terms of studying space weathering, it is
clear that the Gibbs cluster deserves further observations in the
near future. In particular, we encourage observations aimed at developing
better physical characterizations of the members of the Gibbs
cluster, either via broadband photometry or
spectroscopy. In this regard, high quality reflectance spectroscopy of
the largest member, asteroid (20674) 1999~VT$_{1}$, would be extremely
valuable.  A good opportunity to obtain such spectroscopy
will be in September 2014, during the asteroid's next opposition.

Currently available SDSS data for two Gibbs cluster members suggest
that $1.5$ Myr is an insufficient length of time at these heliocentric distances
to transform ordinary anchondrite spectra into more typical $S$-type spectra.
As we have already cautioned though, these conclusions are based on
uncertain data and should be considered preliminary.
Fortunately, uncertainties in the physical properties of the
cluster members has only limited influence our determination of the age of the cluster.
However, as the physical properties of the cluster affect our estimates
of the size of the parent body, better physical characterizations of cluster
members would certainly be useful for refining our understanding of the initial
family forming event.

As explained in Section~\ref{s:obs}, we were unsuccessful in our attempts
to detect comet-like activity among other members of the Gibbs cluster. 
This can be interpreted either as a consequence of the fact that some 
faint activity could actually be present, but it was below the detection 
limits of our observations. The other possibility is that activity 
was actually absent, at least at the times when our observations were carried out.
Activity is known to be transient even for currently active MBCs
\citep[e.g.,][]{hsieh2010,hsieh2011}, and only three of our targets were observed
within the approximate true anomaly range ($-50^{\circ}\lesssim\nu\lesssim90^{\circ}$;
i.e., close to and following perihelion)
where other MBCs have shown activity in the past \citep{hsieh2012}.

However, perhaps the most important issues to keep in mind when interpreting
the lack of any detected activity in our observations of fellow cluster
members of P/2012~F5, is their composition as well as the active nature of
P/2012~F5 itself.
The MBCs are expected to be low-albedo, icy-bearing objects, with spectra most
closely resembling that of $C$-type asteroids.
If it turns out that the Gibbs cluster is 
indeed composed of $Q$-type objects, as suggested by SDSS data, this 
would explain why we did not find any activity in these asteroids.  This hypothesis
is supported by the studies by \citet{stevenson2012} and \citet{moreno2012} who
found that P/2012 F5's activity was most likely due to an impact from another
asteroid and not comet-like sublimation of volatile ices.

Thus, while we did not successfully detect any new comet among the Gibbs cluster,
this result does not necessarily invalidate the hypothesis that young asteroid families and main-belt comets are
linked, as there are several plausible explanations for why no activity was detected.
In fact, if P/2012~F5's activity was due to an impact and not sublimation, making it
a disrupted asteroid and not a comet, we perhaps would not even expect to find other
similar instances of comet-like activity given the low
likelihood of impacts in the asteroid belt, even in families as young as the Gibbs
cluster.  As such, while the Gibbs cluster will certainly be an interesting subject
for further studies of space weathering and catastrophic collisions in the main
asteroid belt, confirmation of the hypothesized link between young families and 
main-belt comets will likely have to come from elsewhere.

\section*{Acknowledgments}
We would like to express our gratitude to Mira Bro\v z and Ricardo Gil-Hutton,
the referees, for the useful comments and suggestions which
helped us to improve this article
The work of B.N. has been supported by the Ministry of
Education and Science of Serbia under the Project 176011,
and , in part, by the European Science Foundation
under the GREAT ESF RNP programme (Exchange Grant No. 3535).
H.H.H. is supported by NASA through Hubble Fellowship grant
HF-51274.01 awarded by the Space Telescope Science Institute,
which is operated by the Association of Universities
for Research in Astronomy (AURA) for NASA, under contract NAS 5-26555
Based in part on observations made with the Italian Telescopio Nazionale Galileo (TNG)
operated on the island of La Palma by the Fundación Galileo Galilei of the INAF
(Istituto Nazionale di Astrofisica) at the Spanish Observatorio del Roque de los
Muchachos of the Instituto de Astrofisica de Canarias. Based in part on observations 
with the University of Hawaii $2.2$~m telescope at Mauna Kea Observatory, 
Institute for Astronomy, University of Hawaii.

\end{document}